\pdfoutput=1

\documentclass[journal]{IEEEtran}
\usepackage {graphicx,alltt,epsf}
\usepackage{times, amsmath, amssymb, epsfig}
\usepackage{bm}
\usepackage[hyphens]{url}
\usepackage[T1]{fontenc}
\usepackage{multirow}
\usepackage{subfigure}

\begin{document}
%\title{Using the Style File IEEEtran.sty}
\title{$L$-infinity Norm Design of Linear-phase Robust Broadband Beamformers using Constrained Optimization} %!PN
\author{R.~C.~Nongpiur,~\IEEEmembership{Member,~IEEE,}
        and~D.~J.~Shpak,~\IEEEmembership{Senior Member,~IEEE} % stops a space
\thanks{Copyright (c) 2012 IEEE. Personal use of this material is permitted. However, permission to use this material for any other purposes must be obtained from the IEEE by sending an email to pubs-permissions@ieee.org.}  
\thanks{R.~C.~Nongpiur and D.~J.~Shpak are with the Department
of Electrical and Computer Engineering, University of Victoria, Victoria,
BC, Canada V8W 3P6 e-mail: rnongpiu@ece.uvic.ca; dshpak@ece.uvic.ca}% <-this % stops a space
\thanks{Manuscript submitted Nov 2012.}}

\maketitle
\begin{abstract}
A new method for the design of linear-phase robust far-field broadband beamformers using constrained optimization is proposed. In the method, the maximum passband\footnote{In this paper, unless explicitly stated, the terms {\it passband} and {\it stopband} refer to the {\it angular passband} and {\it angular stopband} of the beamformer, respectively.} ripple and minimum stopband attenuation are ensured to be within prescribed levels, while at the same time maintaining a good linear-phase characteristic at a prescribed group delay in the passband. Since the beamformer is intended primarily for small-sized microphone arrays where the microphone spacing is small relative to the wavelength at low frequencies, the beamformer can become highly sensitive to spatial white noise and array imperfections if a direct minimization of the error is performed. Therefore, to limit the sensitivity of the beamformer the optimization is carried out by constraining a sensitivity parameter, namely, the white noise gain (WNG) to be above prescribed levels across the frequency band. Two novel design variants have been developed. The first variant is formulated as a convex optimization problem where the maximum error in the passband is minimized, while the second variant is formulated as an iterative optimization problem and has the advantage of significantly improving the linear-phase characteristics of the beamformer under any prescribed group delay or linear-array configuration. In the second variant, the passband group-delay deviation is minimized while ensuring that the maximum passband ripple and stopband attenuation are within prescribed levels. To reduce the computational effort in carrying out the optimization, a nonuniform variable sampling approach over the frequency and angular dimensions is used to compute the required parameters. Experiment results show that beamformers designed using the proposed methods have much smaller passband group-delay deviation for similar passband ripple and stopband attenuation than a modified version of an existing method. 
\end{abstract}

%\vspace{-0.1in}
\section{Introduction}
Microphone arrays are widely used in speech communication applications such as hands-free telephony, hearing aids, speech recognition, and teleconferencing systems. Beamforming is often used with microphone arrays to enhance a speech signal from a preferred spatial direction~\cite{wardBook}. In general, the beamforming approach can be fixed or adaptive, depending upon whether the spatial directivity pattern is fixed~\cite{ward1}-\cite{mabande2}, or varies adaptively on the basis of incoming data~\cite{affes}-\cite{souden}. Though adaptive beamforming performs better when the acoustic environment is time-varying, fixed beamforming is preferred in applications where the direction of the sound source is fixed, such as in in-car communication systems~\cite{gerhardBook} or in hearing aids. In addition, fixed beamformers have lower computational complexity and are easier to implement. 

In many beamformer applications, such as in-car communication systems, voice recognition systems, video conferencing systems, etc., there is often a need to ensure that the gain across the passband has little variation from unity while that in the stopband is always below a prescribed level. At the same time, a passband with good linear-phase characteristics is usually preferred to avoid any signal distortion. Consequently, a straightforward approach for the design of such beamformers is to formulate the problem in terms of the $L_\infty$ norm as it leads to a minimax optimization of the appropriate error functions~\cite{wsluBook}.

In~\cite{ward1}-\cite{doclo1}, designs for broadband beamformers that are not constrained by the size of the array aperture or are based on the assumption of ideal or known microphone characteristics have been proposed. However, in certain applications such as in hearing aid and in-car communication systems there are physical constraints on the array aperture size such that the wavelength of the signal in the lower end of the frequency band is much longer than the maximum allowed aperture length. Consequently, as evident from earlier designs for superdirective narrowband arrays~\cite{cox}-\cite{dale2},  broadband beamformers designed for physically-compact applications can become very sensitive to errors in array imperfections and therefore robustness constraints need to be incorporated in the design. In~\cite{doclo2}-\cite{crocco2}, the statistics of microphone characteristics are taken into account to derive broadband beamformers that are robust to microphone mismatches, while in~\cite{mabande} the white noise gain (WNG) is incorporated in the design to ensure that the beamformer is robust to spatial white noise and array imperfections. The use of the WNG constraint is not new and has been used in earlier beamformer designs to ensure robustness in superdirective beamformers~\cite{cox}-\cite{bitzer}. Interestingly, it has been shown in~\cite{chen2} that the use of statistical properties of the microphone characteristics in~\cite{doclo2} to improve the beamformer robustness  is also a class of the WNG constrained-based technique. Nevertheless, efforts to design robust broadband beamformers with a frequency-invariant beampattern have been carried out only in recent years~\cite{doclo2}-\cite{mabande}. 

In the method in~\cite{mabande2}, the beamformer is designed by performing an $L_2$ minimization of the desired beamformer response while constraining the response at the centre of the passband to unity and the WNG to be above a prescribed level. The method in~\cite{mabande2}, however, is not very effective for designing broadband beamformers where the maximum passband ripple and minimum stopband attenuation is specified since it is based on optimizing the $L_2$ norm of the error rather than the $L_\infty$ norm. Further, the method is only applicable when the prescribed group delay is zero. 

In this paper, we develop a method for designing robust broadband beamformers with good linear-phase characteristics in the passband while ensuring that the maximum passband ripple and minimum stopband attenuation are below and above specified levels, respectively. In the method, we use the $L_\infty$ norm of the error to formulate the optimization problem. Two novel variants have been developed where the WNG is constrained to be above prescribed levels across the frequency band. In the first variant, the beamformer is formulated as a convex optimization problem where the maximum passband error is minimized. The second variant is formulated as an iterative second-order cone programming (SOCP) problem that minimizes the passband group-delay deviation, with the advantage of significantly improving the linear-phase characteristics of the beamformer under any prescribed group delay or linear-array configuration. The iterative SOCP has been extended from the design of IIR filters, such as in~\cite{hinamoto} and~\cite{nongpiur}, to optimize the non-linear formulations of the WNG and the group-delay deviation in the proposed method. It should be noted that in~\cite{tsui}, conformal arrays were designed by formulating the optimization problem as an iterative SOCP; however, the method in~\cite{tsui} is primarily confined to the design of narrowband arrays where only the magnitude of the desired beam pattern is specified. 

To reduce the high computational effort associated with designing broadband beamformers using the iterative SOCP method, we extended the nonuniform sampling method, developed in~\cite{antoniou2} for the design of digital filters, to two dimensions: frequency and angles; this has resulted in a reduction of the computational effort by more than an order of magnitude for the SOCP method. Experimental results show that beamformers designed using the proposed methods have superior performance when compared to a modified version of an existing method. 

The paper is organized as follows. In Section II, we describe the filter-and-sum beamformer and the associated formulations of the beamformer response and WNG for a uniform linear array in the far field. Then in Section III, we develop formulations for variant 1 of the proposed method where the design is formulated as a convex optimization problem. In Section IV, we develop formulations for variant 2 of the proposed method where the design is formulated as an iterative SOCP problem. Then in Section V, performance comparisons between the proposed design variants with a modified version of an existing method are carried out. Conclusions are drawn in Section VI.
\vspace{-0.1in}
\section{Far-field Broadband Beamforming}
\begin{figure}
\begin{center}
\includegraphics[width=0.45\textwidth]{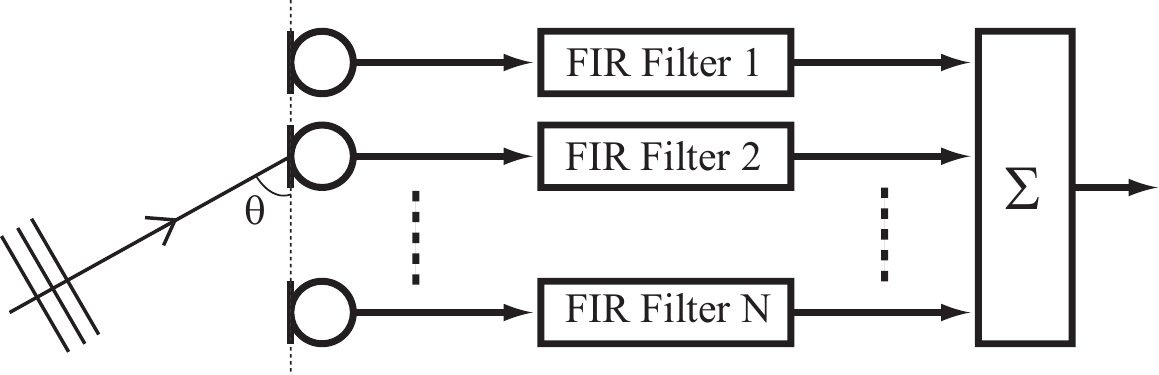}
\caption{Filter and sum broadband beamformer.}
\label{filAndSumBf}
\end{center}
%\vspace{-0.3in}
\end{figure}
In this paper, we assume a far-field signal impinging on a linear microphone array that is realized as a filter-and-sum beamformer, as shown in Fig.~\ref{filAndSumBf}. The microphones are assumed to be omnidirectional and the filters are FIR. If $N$ is the number of microphones and $L$ is the length of each filter, the response of the filter-and-sum beamformer is given by~\cite{wardBook}
\begin{equation}
B(\omega, \theta) = \sum_{n=0}^{N-1} \sum_{l=0}^{L-1} x_{n, l} g_{n, l}(\omega, \theta)
\label{bmResp}
\end{equation}
where
\begin{equation}
g_{n, l}(\omega, \theta) = \exp \left[ -j\omega \left(\frac{f_s d_n \cos\theta}{c} + l \right) \right]
\end{equation}
$\omega$ is the frequency in radians, $\theta$ is the direction of arrival, $c$ is the speed of sound in air, $f_s$ is the sampling frequency, $d_n$ is the distance of the $n$th microphone from the origin, and $x_{n, l}$ is the $l$th coefficient of $n$th FIR filter. In matrix form, (\ref{bmResp}) can be expressed as
\begin{equation}
B(\mathbf{x}, \omega, \theta) = \sum_{n=0}^{N-1}  \mathbf{g}_n(\omega, \theta)^T \mathbf{x}_n = \mathbf{g}(\omega, \theta)^T \mathbf{x}
\label{bmResp2}
\end{equation}
where
\begin{eqnarray}
\mathbf{x}^T & = & \left[\mathbf{x}_0^T ~\mathbf{x}_1^T \cdots ~\mathbf{x}_{N-1}^T \right]  \\
 \mathbf{g}(\omega, \theta)^T & = & \left[ \mathbf{g}_0(\omega, \theta)^T ~\mathbf{g}_1(\omega, \theta)^T \cdots ~\mathbf{g}_{N-1}(\omega, \theta)^T \right]  \\
\mathbf{x}_n & = & \left[ x_{n, 0} ~x_{n, 1} \cdots ~x_{n, L-1} \right]^T \label{xElements} \\
\mathbf{g}_n(\omega, \theta) & = & \left[ g_{n, 0}(\omega, \theta) ~g_{n, 1}(\omega, \theta) \cdots ~g_{n, L-1}(\omega, \theta) \right]^T  
\end{eqnarray}
If $\theta_d$ is the desired steering angle of the beamformer, the WNG of the beamformer is given by~\cite{wardBook}
\begin{equation}
G_w(\mathbf{x}, \omega)  =   \frac{|B(\mathbf{x}, \omega, \theta_d)|^2}{ \displaystyle \sum_{n=0}^{N-1} \left| \sum_{l=0}^{L-1} x_{n, l} e^{-j\omega l}\right|^2 }  = \frac{|\mathbf{g}(\omega, \theta_d)^T \mathbf{x}|^2}{\|\mathbf{A}(\omega) \mathbf{x}\|_2^2} \label{wng}
\end{equation}
where
\begin{eqnarray}
\mathbf{A}(\omega) & = & \mathbf{I}_N \otimes \mathbf{a}(\omega)^T \\
\mathbf{a}(\omega)^T & = & \left[ 1~ e^{-j\omega}~\cdots ~e^{-j(L-1)\omega} \right]^T \notag \\
\end{eqnarray}
$\mathbf{I}_N$ is an $N\times N$ identity matrix, $\otimes$ is the Kronecker product, and $\|\mathbf{v}\|_2$ is the $L_2$ norm of vector $\mathbf{v}$.

\section{Beamformer Design as a Convex Optimization Problem}
If $B_d(\omega, \theta)$ is the desired beampattern at all frequencies and directions, the error between the beamformer response and the desired beampattern is given by
\begin{equation}
e_b(\mathbf{x}, \omega, \theta) = B(\mathbf{x}, \omega, \theta) - B_d(\omega, \theta)
\label{beamError}
\end{equation}
The $L_p$ norm of the error across the passband directions $\Theta_{pb} \in [\theta_{pl}, \theta_{ph}]$ and frequency band of interest $\Omega \in [\omega_l, \omega_h]$ is given by
\begin{eqnarray}
E^{(pb)}_p(\mathbf{x}) & = & \left[ \int_{\Omega} \int_{\Theta_{pb}} | e_b(\mathbf{x}, \omega, \theta) |^p d\theta~d\omega \right]^{1/p} \notag \\
& \approx & \kappa_p \left[ \sum_{m=1}^M \sum_{k=1}^{K_{pb}}  |e_b(\mathbf{x}, \omega_m, \theta^{(pb)}_k)|^p \right]^{1/p} \notag \\
& = & \| \mathbf{U}_{pb}\mathbf{x} - \mathbf{d}_{pb} \|_p
\label{L2PbBeamError}
\end{eqnarray}
where
\begin{eqnarray}
\mathbf{U}_{pb} & = & \left[ \kappa_p\mathbf{g}(\omega_1, \theta^{(pb)}_1) \cdots \kappa_p\mathbf{g}(\omega_1, \theta^{(pb)}_K) \right.\cdots \notag \\
& & \ \ \ \ \ \ \ \ \left. \kappa_p\mathbf{g}(\omega_M, \theta^{(pb)}_1) \cdots \kappa_p\mathbf{g}(\omega_M, \theta^{(pb)}_K) \right]^T \label{BpbMat} 
\end{eqnarray}
\begin{eqnarray}
\mathbf{d}_{pb} & = & \left[\kappa_p B_d(\omega_1, \theta^{(pb)}_1) \cdots \kappa_p B_d(\omega_1, \theta^{(pb)}_K) \right. \cdots \notag \\
& & \ \ \ \ \left. \kappa_p B_d(\omega_M, \theta^{(pb)}_1) \cdots \kappa_p B_d(\omega_M, \theta^{(pb)}_K)  \right]^T  
\label{dpbDefn} 
\end{eqnarray}
$\theta^{(pb)}_k \in \Theta_{pb}$, $\omega_m \in \Omega$, and $\kappa_p$ is a constant. For the stopband region, defined by $\Theta_{sb} \in [\theta_{sl}, \theta_{sh}]$, we set $B_d(\omega, \theta) = 0$ and, as a consequence, the $L_p$ norm of the stopband error is given by
\begin{eqnarray}
E^{(sb)}_p(\mathbf{x}) & = & \left[ \int_{\Omega} \int_{\Theta_{sb}} | e_b(\mathbf{x}, \omega, \theta) |^p d\theta~d\omega \right]^{1/p} \notag \\
& \approx & \kappa_s \left[ \sum_{m=1}^M \sum_{k=1}^{K_{sb}}  |e_b(\mathbf{x}, \omega_m, \theta^{(sb)}_k)|^p \right]^{1/p} \notag \\
& = & \| \mathbf{U}_{sb}\mathbf{x} \|_p
\label{L2SbBeamError}
\end{eqnarray}
where
\begin{eqnarray}
\mathbf{U}_{sb} & = & \left[ \kappa_s\mathbf{g}(\omega_1, \theta^{(sb)}_1) \cdots \kappa_s\mathbf{g}(\omega_1, \theta^{(sb)}_K) \right.\cdots \notag \\
& & \ \ \ \ \ \ \ \ \left. \kappa_s\mathbf{g}(\omega_M, \theta^{(sb)}_1) \cdots \kappa_s\mathbf{g}(\omega_M, \theta^{(sb)}_K) \right]^T \label{BsbMat}
\end{eqnarray}
$\theta^{(sb)}_k \in \Theta_{sb}$ and $\kappa_s$ is a constant. To obtain the filter coefficients for a broadband beamformer, the optimization problem is solved by minimizing the passband error, $E^{(pb)}_p(\mathbf{x})$, while constraining the stopband error to be below a prescribed threshold. To ensure robustness, the WNG is also constrained to be above prescribed levels across the frequency band; that is
\begin{eqnarray}
\mbox{minimize } & &  E^{(pb)}_p(\mathbf{x})  \label{optProblem1}   \\
\mbox{subject to: } & & E^{(sb)}_p(\mathbf{x}) \leq \Gamma_{sb} \notag \\
 & & G_w(\mathbf{x}, \omega_m) \geq \Gamma_{wng}(\omega) \ \ \forall \  \ \omega_m \in \Omega \notag 
\end{eqnarray}
where $\Gamma_{sb}$ is the minimum stopband attenuation and $\Gamma_{wng}(\omega)$ is the minimum WNG at frequency $\omega$. Since the WNG constraint in (\ref{optProblem1}) is a non-linear constraint that is non-convex, the optimization problem in (\ref{optProblem1}) is accordingly non-convex. 

For a linear-phase beamformer response with prescribed group delay of $\tau$, the desired passband beamformer response, $B_d(\omega, \theta)$, is given by
\begin{equation}
B_d(\omega, \theta) = e^{-j\omega\tau_d}
\label{bfLinResp}
\end{equation}
Noting that 
\begin{equation}
|B(\mathbf{x}, \omega, \theta_d)| = |e^{j\omega\tau_d} B(\mathbf{x}, \omega, \theta_d)| \geq |\Re [e^{j\omega\tau_d} B(\mathbf{x}, \omega, \theta_d)]|
\end{equation}
it is clear that the WNG in (\ref{wng}) will satisfy
\begin{equation}
G_w(\mathbf{x}, \omega) = \frac{|B(\mathbf{x}, \omega, \theta_d)|^2}{ \displaystyle \sum_{n=0}^{N-1} \left| \sum_{l=0}^{L-1} x_{n, l} e^{-j\omega l}\right|^2 } \geq \frac{|\Re [e^{j\omega\tau_d} B(\mathbf{x}, \omega, \theta_d)]|^2}{ \displaystyle \sum_{n=0}^{N-1} \left| \sum_{l=0}^{L-1} x_{n, l} e^{-j\omega l}\right|^2 } 
\label{wngInequality}
\end{equation}
Therefore, if 
\begin{equation}
\frac{|\Re [e^{j\omega\tau_d} B(\mathbf{x}, \omega, \theta_d)]|^2}{ \displaystyle \sum_{n=0}^{N-1} \left| \sum_{l=0}^{L-1} x_{n, l} e^{-j\omega l}\right|^2 }  \geq \Gamma_{wng}
\label{wngAproxIneq}
\end{equation}
then the condition $G_w(\mathbf{x}, \omega) > \Gamma_{wng}(\omega)$ will always be true. Taking the square root on both sides of (\ref{wngAproxIneq}), substituting from (\ref{wng}), and doing some simple algebraic manipulation we get
\begin{equation}
\sqrt{\Gamma_{wng}} \   \|\mathbf{A}(\omega) \mathbf{x}\|_2 - \left| \Re [e^{j\omega\tau_d} B(\mathbf{x}, \omega, \theta_d)]\right|   \leq  0
\label{wngAproxIneq2}
\end{equation}
If the desired response in the passband is linear phase as in (\ref{bfLinResp}), it is clear that the minimization of $|B(\mathbf{x}, \omega, \theta_d) - e^{-j\omega\tau_d}|$ will result in a beamformer solution where the term, $e^{j\omega\tau_d}B(\mathbf{x}, \omega, \theta_d)$, is approximately unity with an imaginary component that is very small compared to unity. This implies that
\begin{equation}
\begin{split}
|B(\mathbf{x}, \omega, \theta_d)| & = |e^{j\omega\tau_d} B(\mathbf{x}, \omega, \theta_d)| \approx \\
& |\Re [e^{j\omega\tau_d} B(\mathbf{x}, \omega, \theta_d)]| = \Re [e^{j\omega\tau_d} B(\mathbf{x}, \omega, \theta_d)] \approx 1
\end{split}
\label{respApprox}
\end{equation}
which further implies that the inequality in (\ref{wngInequality}) is approximately an equality, thereby making the inequalities in (\ref{wngAproxIneq}) and (\ref{wngAproxIneq2}) approximately equivalent to the WNG inequality constraint in (\ref{optProblem1}). Using (\ref{wng}) and (\ref{respApprox}) the inequality in (\ref{wngAproxIneq2}) can be expressed as 
\begin{equation}
\sqrt{\Gamma_{wng}} \   \|\mathbf{A}(\omega) \mathbf{x}\|_2 -  \Re [e^{j\omega\tau_d} \mathbf{g}(\theta_d, \omega)^T \mathbf{x}]   \leq  0
\label{wngAproxIneq3}
\end{equation}
which turns out to be a convex inequality. To ensure that the design results in a minimization of the maximum passband ripple while the maximum stopband gain is below $\Gamma_{sb}$, the $L_\infty$ norm of the error is optimized; that is, we set $p = \infty$ in (\ref{L2PbBeamError}) and (\ref{L2SbBeamError}). With these modifications, and using the convex WNG constraint in (\ref{wngAproxIneq3}), the problem in (\ref{optProblem1}) can now be formulated as a convex optimization problem given by
\begin{eqnarray}
\mbox{minimize } & & \| \mathbf{U}_{pb}\mathbf{x} - \mathbf{d}_{pb}^{(lin)} \|_\infty \label{optProblem4}   \\
\mbox{subject to: } & & \| \mathbf{U}_{sb}\mathbf{x} \|_\infty \leq \Gamma_{sb} \notag \\
 & & \sqrt{\Gamma_{wng}} \   \|\mathbf{A}(\omega_m) \mathbf{x}\|_2 - \notag \\
 & & \ \ \ \ \ \ \Re [e^{j\omega_m\tau_d}\mathbf{g}(\theta_d, \omega_m)^T \mathbf{x}]  \leq  0 \ \ \forall \ \omega_m \in \Omega \notag 
\end{eqnarray}
where $\mathbf{x} \in \mathbf{R}^{NL}$ is the optimization variable and $\mathbf{d}_{pb}^{(lin)} \in \mathbf{R}^{MK}$ is the desired-response vector in (\ref{dpbDefn}) when $B_d(\omega, \theta)$ has a linear-phase response as in (\ref{bfLinResp}).

\subsection{Special case of symmetric and perfectly linear-phase beamformers for symmetric microphone arrays}
If the beampattern is symmetric about $\theta = \pi/2$ in magnitude and phase, and the positions of the array sensors are symmetric with respect to the array center such that 
\begin{equation}
d_{N-n-1} = -d_n
\label{symCondt0}
\end{equation}
then the filters of the beamformer will satisfy the symmetry condition
\begin{equation}
x_{n, l} = x_{N-n-1, l}
\label{symCondt}
\end{equation}
Conversely, if the position of the array sensors are symmetric with respect to the array center and condition (\ref{symCondt}) is satisfied, then the beampattern is always symmetric. Since (\ref{symCondt}) is an affine condition, it can therefore be incorporated in the convex optimization problem in (\ref{optProblem4}) as an additional constraint.

If the desired group delay is set to 
\begin{equation}
\tau_{hlf} = (L-1)f_s^{-1}/2
\label{tauSym0}
\end{equation}
then the beampattern is guaranteed to be perfectly linear phase if the condition 
\begin{equation}
x_{n, l} = x_{N-n-1, L-l-1}
\label{linPhaseConstr}
\end{equation}
is satisfied, in addition to the condition in (\ref{symCondt0}). Note that the linear-phase condition in (\ref{linPhaseConstr}) is applicable even for non-symmetric beamformers.

The symmetry and linear-phase conditions lead to a simplification of the beamformer response where the number of variables is approximately reduced by a factor of about 2 as shown in Appendixes A and B. Therefore, using these simplifications, it is indeed possible to reformulate the optimization problem so that the number of variables in the optimization problem is reduced by about a factor of 2. Note that for the special case where the desired beampattern is symmetric about $\theta = \pi/2$ and the desired group delay is $\tau_{hlf}$, the number of variables can be reduced even further, by a factor of 4, as shown in~\cite{crocco2}.

\section{Beamformer Design as an Iterative  Problem}
In this section, the objective is to minimize the passband group-delay deviation while ensuring that the maximum passband ripple and minimum stopband attenuation are within prescribed specifications and the WNG is above prescribed levels across the frequency band. Since the group-delay deviation, the passband response error, and the WNG are non-linear functions that are non-convex, we frame the optimization as an iterative constrained optimization problem by approximating each update as a linear approximation step as in~\cite{hinamoto}. To this end, we derive formulations for the group-delay deviation, the passband response error, and the white noise gain. Then, we incorporate the formulations within the framework of a constrained optimization problem.
\subsection{Group-delay deviation}
The group delay of the beamformer response, $B(\omega, \theta)$, is given by
\begin{equation}
\tau(\omega, \theta) = -\frac{d\theta_B}{d\omega}
\label{tauDefn}
\end{equation}
where 
\begin{equation}
\theta_B = \arg B(\omega, \theta)
\end{equation}
From Appendix C, it is easy to show that the group delay of the beamformer simplifies to 
\begin{equation}
\tau(\mathbf{x}, \omega, \theta) = -\frac{\alpha_1(\mathbf{x}, \omega, \theta)\alpha_2(\mathbf{x}, \omega, \theta)+\beta_1(\mathbf{x}, \omega, \theta)\beta_2(\mathbf{x}, \omega, \theta) }{\alpha_1(\mathbf{x}, \omega, \theta)^2 + \beta_1(\mathbf{x}, \omega, \theta)^2 }
\label{grpDelSimple}
\end{equation}
where
\begin{eqnarray}
\alpha_1(\mathbf{x}, \omega, \theta) & = & \sum_{n=0}^{N-1} \sum_{l=0}^{L-1} x_{nl} \cos (\omega k_{nl}) \\
\alpha_2(\mathbf{x}, \omega, \theta) & = & \sum_{n=0}^{N-1} \sum_{l=0}^{L-1} x_{nl} k_{nl} \cos (\omega k_{nl}) 
\end{eqnarray}
\begin{eqnarray}
\beta_1(\mathbf{x}, \omega, \theta) & = & \sum_{n=0}^{N-1} \sum_{l=0}^{L-1} x_{nl} \sin (\omega k_{nl}) \\
\beta_2(\mathbf{x}, \omega, \theta) & = & \sum_{n=0}^{N-1} \sum_{l=0}^{L-1} x_{nl} k_{nl} \sin (\omega k_{nl}) 
\end{eqnarray}
\begin{eqnarray}
k_{nl} & = & -\frac{f_s d_n \cos \theta}{c} - l
\end{eqnarray}
The group-delay error at frequency $\omega$ is given by
\begin{equation}
e_g(\mathbf{x}, \omega, \theta) = \tau(\mathbf{x}, \omega, \theta) - \tau_d
\end{equation}
where $\tau_d$ is the prescribed group delay. If $\mathbf{x}_k$ is the value of $\mathbf{x}$ at the start of the $k$th iteration and ${\bm \delta}$ is the update to $\mathbf{x}_k$, the updated value of the group-delay error can be estimated by a linear approximation
\begin{equation}
e_g(\mathbf{x}_k+{\bm \delta}, \omega, \theta) \approx e_g(\mathbf{x}_k, \omega, \theta) + \nabla e_g(\mathbf{x}_k, \omega, \theta)^T {\bm \delta}
\label{gdError}
\end{equation}
which becomes more accurate as $\|{\bm \delta} \|_2$ gets smaller.

The $L_p$-norm of the passband group-delay error for the $k$th iteration is given by
\begin{eqnarray}
\mathcal{E}_p^{(gd)}(\mathbf{x}_k) & = &  \left[ \int_{\Omega} \int_{\Theta_{pb}} | e_g(\mathbf{x}_{k+1}, \omega, \theta)|^p d\theta d\omega \right]^{1/p} \notag \\
& \approx & \kappa_{g} \left[ \sum_{m=1}^M \sum_{n=1}^{K_{gd}} |e_g(\mathbf{x}_{k+1}, \omega_m, \theta_n)|^p \right]^{1/p} \notag \\
& \approx & \left[ \sum_{m=1}^M \sum_{n=1}^{K_{gd}} |\kappa_{g} e_g(\mathbf{x}_k, \omega_m, \theta_n) + \right. \notag \\
& &  \ \ \ \ \kappa_{g} \nabla e_g(\mathbf{x}_k, \omega_m, \theta_n)^T {\bm \delta}|^p \Big]^{1/p} \notag \\
\label{gdDev}
\end{eqnarray}
where $\omega_m \in \Omega$, $\theta_n \in \Theta_{pb}$, and $\kappa_{g}$ is a constant. Expressing (\ref{gdDev}) in matrix form, we get
\begin{eqnarray}
\mathcal{E}_p^{(gd)}(\mathbf{x}_k) & \approx & \|\mathbf{C}_k {\bm \delta} + \mathbf{d}_k \|_p  \label{gdDevMat}
\end{eqnarray}
where
\begin{eqnarray}
\mathbf{C}_k &=& \left[ \begin{matrix}
\kappa_{g} \nabla e_g(\mathbf{x}_k, \omega_1, \theta_1)^T  \cr
\vdots   \cr
\kappa_{g} \nabla e_g(\mathbf{x}_k, \omega_M, \theta_K)^T  \cr
\end{matrix}
\right ]  \\
\mathbf{d}_k &=& [d_{11}~d_{12}~\cdots~d_{MK}]^T, \\
d_{mn} & = & \kappa_{g} e_g(\mathbf{x}_k, \omega_m, \theta_n), \ \  \omega_m \in \Omega, \theta_n \in \Theta_{pb}
\label{CmatParam}
\end{eqnarray}
The right-hand side of (\ref{gdDevMat}) is the $L_p$-norm of an affine function of ${\bm \delta}$ and, therefore, it is convex with respect to ${\bm \delta}$~\cite{wsluBook}.

\subsection{Passband Response Error}
Since the minimization of the passband group-delay deviation  results in a linearization of the phase, we compute the passband response error by considering only the magnitude response so as to facilitate greater decoupling between the passband response error and the group delay deviation. This is because greater decoupling or independence between the optimization parameter results in greater degrees of freedom and, in turn, a better solution. Consequently, the passband response error is given by
\begin{equation}
e_r(\mathbf{x}, \omega, \theta) = |B(\mathbf{x}, \omega, \theta)|^2 - |B_d(\omega, \theta)|^2
\label{magRespError}
\end{equation}
where $\omega \in \Omega$ and $\theta \in \Theta_{pb}$. Using the same approach as in Subsection IV-A, the $L_p$-norm of the passband response error, $\mathcal{E}_p^{(pb)}(\mathbf{x}_k)$, in matrix form can be approximated as
\begin{eqnarray}
\mathcal{E}_p^{(pb)}(\mathbf{x}_k) & \approx & \|\mathbf{D}_k {\bm \delta} + \mathbf{f}_k \|_p  \label{pbErrMat}
\end{eqnarray}
where
\begin{eqnarray}
\mathbf{D}_k &=& \left[ \begin{matrix}
\kappa_{r} \nabla e_r(\mathbf{x}_k, \omega_1, \theta_1)^T  \cr
\vdots   \cr
\kappa_{r} \nabla e_r(\mathbf{x}_k, \omega_M, \theta_K)^T  \cr
\end{matrix}
\right ] \label{DmatParam} \\
\mathbf{f}_k &=& [f_{11}~f_{12}~\cdots~f_{MK}]^T, \\
f_{mn} & = & \kappa_{r} e_r(\mathbf{x}_k, \omega_m, \theta_n), \ \  \omega_m \in \Omega, \theta_n \in \Theta_{pb}
\end{eqnarray}
\subsection{White Noise Gain}
If $\Gamma_{wng}(\omega)$ is the prescribed lower bound of the WNG at frequency $\omega$, the difference between the WNG of the beamformer and the prescribed lower bound is given by
\begin{equation}
e_w(\mathbf{x}, \omega)  = G_w(\mathbf{x}, \omega) - \Gamma_{wng}(\omega)
\end{equation}
As in Subsection IV-A, the update of $e_w(\mathbf{x}, \omega)$ for the $k$th iteration can be approximated as
\begin{equation}
e_w(\mathbf{x}_k+{\bm \delta}, \omega) \approx e_w(\mathbf{x}_k, \omega) + \nabla e_w(\mathbf{x}_k, \omega)^T{\bm \delta}, \ \ \ \omega \in \Omega
\label{wngDiffUpdate}
\end{equation}
Sampling the RHS of (\ref{wngDiffUpdate}) across $\Omega$ and expressing it in matrix form we get
\begin{equation}
\mathbf{w}(\mathbf{x}_k) = \mathbf{Q}_k{\bm \delta} + \mathbf{h}_k
\label{wngDiffUpdateMat}
\end{equation}
where
\begin{eqnarray}
\mathbf{Q}_k &=& \left[ \begin{matrix}
\nabla e_w(\mathbf{x}_k, \omega_1)^T \cr
\vdots   \cr
\nabla e_w(\mathbf{x}_k, \omega_M)^T \cr
\end{matrix}
\right ]  \label{QvalXX} \\
\mathbf{h}_k &=& [e_w(\mathbf{x}_k, \omega_1)~\cdots~e_w(\mathbf{x}_k, \omega_M)]^T, \label{fvalXX}
\end{eqnarray}
and $\omega_m \in \Omega$.
\vspace{-0.05in}
\subsection{Formulating the Optimization Problem}
The optimization problem is solved using a two-step method. The objective of the first step is to obtain a good starting point for the second step. In the first step, the passband error is minimized under the constraint that the stopband error is below a prescribed threshold and the WNG is above prescribed levels across the frequency spectrum. In the second step, we instead minimize the passband group-delay deviation while constraining the stopband error and the WNG as in the first step; additionally, we also constrain the passband error so that it does not exceed that of the beamformer solution obtained in the first step.

For the first step, we consider two initializing beamformers. The first initializing beamformer is obtained by solving the optimization problem in (\ref{optProblem4}). The second is obtained by modifying the problem in (\ref{optProblem4}) to include a regularization term so that the filter coefficients remain small. We found that in many cases, the regularized version facilitates faster convergence and results in better solutions for the second step\footnote{A possible reason is because in the iterative optimization problem there is an $L_2$-norm constraint on the maximum coefficient update thereby resulting in more iterations if the distance between the starting and final solutions is greater, which is often the case when the starting solution is not regularized.}. The modified problem is given by 
\begin{eqnarray}
\mbox{minimize } & & \| \mathbf{U}_{pb}\mathbf{x} - \mathbf{d}_{pb}^{(lin)} \|_\infty + \lambda \|\mathbf{x}\|_2 \label{optProblem4x}   \\
\mbox{subject to: } & & \| \mathbf{U}_{sb}\mathbf{x} \|_\infty \leq \Gamma_{sb} \notag \\
 & & \sqrt{\Gamma_{wng}} \   \|\mathbf{A}(\omega_m) \mathbf{x}\|_2 - \notag \\
 & & \ \ \ \ \ \ \Re [e^{j\omega_m\tau_d}\mathbf{g}(\theta_d, \omega_m)^T \mathbf{x}]  \leq  0 \ \ \forall \ \omega_m \in \Omega \notag 
\end{eqnarray}
where $\lambda$ is a small positive value. In our experiments, $\lambda$ was set to 0.01. It should be noted that though the beamformer solution of (\ref{optProblem4x}) may work better for initialization, it usually has lower performance when compared with that of (\ref{optProblem4}). In all cases, we found that using any of the two initialization solutions always result in a final beamformer solution that has much smaller group delay deviation. Therefore, if there is a need to reduce the amount of computation required, only the second initializing beamformer can be used.

For the second step, the solution is obtained by solving an iterative optimization problem where the group-delay error in (\ref{gdDevMat}) is minimized under the constraint that the passband and stopband errors in (\ref{pbErrMat}) and (\ref{L2SbBeamError}), respectively, are below prescribed thresholds and the WNG in (\ref{wngDiffUpdateMat}) is above prescribed levels across the frequency spectrum; i.e.,
\begin{eqnarray}
\mbox{minimize } & & \mathcal{E}^{(gd)}_p(\mathbf{x}_k)  \label{iterOptProblem1}   \\
\mbox{subject to: } & & \mathbf{w}(\mathbf{x}_k) \geq \mathbf{0} \notag \\
& & \mathcal{E}^{(pb)}_p(\mathbf{x}_k) \leq \Gamma_{pb} \notag \\
& & E^{(sb)}_p(\mathbf{x}_k+{\bm \delta}) \leq \Gamma_{sb} \notag \\
& & \|{\bm \delta}\|_2 \mbox{ is small} \notag
\end{eqnarray}
where $\Gamma_{pb}$ and $\Gamma_{sb}$ are the passband and stopband thresholds, respectively, and $\mathbf{0} \in \mathbf{R}^M$. Note that the errors $\mathcal{E}^{(gd)}_p (\mathbf{x})$ and $\mathcal{E}^{(pb)}_p(\mathbf{x})$ provide the useful flexibility for independently controlling the passband phase characteristics and magnitude response in the optimization. To ensure that the maximum group-delay error is minimized under the constraint that the maximum passband ripple and minimum stopband attenuation are below prescribed specifications, we consider their $L_\infty$-norm, that is, $\mathcal{E}^{(gd)}_\infty$ $\mathcal{E}^{(pb)}_\infty$ and $E^{(sb)}_\infty$, which gives
\begin{eqnarray}
\mbox{minimize } & & \mathcal{E}^{(gd)}_\infty(\mathbf{x}_k)  \label{iterOptProblem2}   \\
\mbox{subject to: } & & \mathbf{w}(\mathbf{x}_k) \geq \mathbf{0} \notag \\
& & \mathcal{E}^{(pb)}_\infty(\mathbf{x}_k) \leq \Gamma_{pb} \notag \\
& & E^{(sb)}_\infty(\mathbf{x}_k+{\bm \delta}) \leq \Gamma_{sb} \notag \\
& & \|{\bm \delta}\|_2 \leq \Gamma_\delta (k) \notag
\end{eqnarray}
where $\Gamma_\delta (k)$ ensures that the $L_2$ norm of the update is small. The threshold for the passband response error is obtained by taking the $L_\infty$ norm of the passband response error of the beamformer solution obtained in the first step as
\begin{equation}
\Gamma_{pb} = \| |\mathbf{U}_{pb}\mathbf{x}_{sol1}|^2 - |\mathbf{d}_{pb}|^2 \|_\infty \pm \epsilon_f
\end{equation}
where $\mathbf{x}_{sol1}$ is the beamformer solution obtained in the first step and $\epsilon_f$ is a small positive value for fine tuning the maximum passband ripple.

During the starting phase of the optimization iterations it is quite possible that the stopband error may not satisfy the prescribed threshold or the WNG constraint may not be satisfied at some of the frequency points. To ensure that the optimization problem does not become infeasible, we relax the two inequality constraints by adding or subtracting the bounds with a slack variable, $\delta_{rlx}$, which is also minimized; when $\delta_{rlx} = 0$, the original constraints are restored. Furthermore, to speed up the convergence, $\Gamma_\delta (k)$ can be made relatively large during the starting iteration and gradually reduced to a small fixed value after a certain number of iterations. With these modifications the optimization problem becomes
\begin{eqnarray}
\mbox{minimize } & & \|\mathbf{C}_k {\bm \delta} + \mathbf{d}_k \|_\infty + W \delta_{rlx} \label{iterOptProblem4}   \\
\mbox{subject to: } & & \mathbf{Q}_k{\bm \delta} + \mathbf{h}_k \geq \mathbf{0} - \delta_{rlx} \notag \\
& & \| \mathbf{D}_k {\bm \delta} + \mathbf{f}_k \|_\infty \leq \Gamma_{pb} + \delta_{rlx}  \notag \\
& & \| \mathbf{U}_{sb}(\mathbf{x}_k+{\bm \delta}) \|_\infty \leq \Gamma_{sb} + \delta_{rlx} \notag \\
& & \|{\bm \delta}\|_2 \leq \Gamma_\delta (k) + \delta_{rlx}  \notag \\
& & \delta_{rlx} \geq 0 \notag
\end{eqnarray}
where ${\bm \delta} \in \mathbf{R}^{LN}$ and $\delta_{rlx} \in \mathbf{R}^{1}$ are the optimization variables and 
\begin{equation}
\Gamma_\delta (k) = \begin{cases} \gamma_k & k < T \\
\gamma_{small} & \mbox{otherwise}
\end{cases}
\end{equation}
such that $\gamma_i > \gamma_{i+1}$ and $W > 0$. Consequently, the 2-step method can be summarized as follows:
\subsubsection*{Step A-1}
Solve the convex optimization problem in (\ref{optProblem4x}).
\subsubsection*{Step A-2}
Solve the iterative algorithm in (\ref{iterOptProblem4}) using the beamformer obtained in {\it A-1} for initialization.
\subsubsection*{Step B-1 (optional)}
Solve the convex optimization problem in (\ref{optProblem4}).
\subsubsection*{Step B-2 (optional)}
Solve the iterative algorithm in (\ref{iterOptProblem4}) using the beamformer obtained in {\it B-1} for initialization.
\subsubsection*{Step C}
If {\it Steps B-1} and {\it B-2} are used, compare the beamformers obtained in {\it Steps A-2} and {\it B-2} and take the one with the smaller group-delay deviation. Otherwise, the solution from {\it Step A-2} is taken.
 
The optional steps, {\it B-1} and {\it B-2}, can be carried out if the amount of computation required is not a critical factor, in order to increase the possibility for obtaining a better solution. Note that in our experiments we have used the optional steps.
 
The optimization problem in (\ref{iterOptProblem4}) can be readily expressed as {\it second order cone programmming} (SOCP) problems as in~\cite{hinamoto} and solved using efficient SOCP solvers such as the one available in the SeDuMi optimization toolbox for MATLAB~\cite{sedumi}.

\subsection{Special case of symmetric FIR filters with symmetric microphone array} 
For the iterative optimization problem, the symmetry constraint in (\ref{symCondt}) can be formulated for the $k$th iteration as
\begin{equation}
x_{n, l}^{(k)}+\delta_{n, l} = x_{N-n-1, l}^{(k)}+\delta_{N-n-1, l}
\label{iterSymConstr}
\end{equation}
where $x_{n, l}^{(k)}$ is an element of the $k$th iteration of $x$ as defined in (\ref{bmResp2}), and $\delta_{n, l}$ the corresponding update for that iteration. Though the symmetry constraint in (\ref{iterSymConstr}) guarantees that the beampattern is perfectly symmetric, it will not always result in the minimum group-delay deviation for the same passband and stopband specifications, since there can exist other solutions that do not satisfy the constraint in (\ref{iterSymConstr}) but have smaller group-delay deviation.

\subsection{Non-uniform Variable Sampling in Frequencies and Angles}
In~\cite{antoniou2}, a non-uniform variable sampling technique for $L_\infty$-norm optimization of digital filters was proposed. The technique was found to be very effective in eliminating the spikes in the error functions while at the same time reducing the computational complexity by an order of magnitude. However, a direct application of the technique in beamformer optimization is not possible since the technique works only in the frequency dimension whereas beamformer optimization requires sampling in the two dimensions of frequency and angle. In this subsection, we extend the technique in~\cite{antoniou2} to two dimensions so that it can be used to solve optimization problems for beamformer design.

In the dimensions of frequency and angle, the extended technique involves the following steps:
\subsubsection*{Step A}
Evaluate the required error function of $\omega$ and $\theta$ with respect to a dense uniform 2-dimension (2-D) grid that spans the frequency band on one side and the angular band on the other, say, $(\bar{\omega}_1, \bar{\theta}_1)$, $\ldots$ $(\bar{\omega}_P, \bar{\theta}_1)$, $\ldots$ $(\bar{\omega}_P, \bar{\theta}_Q)$ where $P$ and $Q$ are fairly large of the order of $10\times M$ and $10\times K$, respectively.
\subsubsection*{Step B} 
Segment the 2-D plane into rectangular blocks such that there are $M$ blocks along the dimension of $\omega$ and $K$ blocks along the dimension of $\theta$.
\subsubsection*{Step C} 
For each of the rectangular blocks find the frequency-angle pair that yields the maximum error. Let the frequency-angle pairs be $(\omega_m, \theta_k)$ where $m=1, 2, \ldots, M$ and $k = 1, 2, \ldots, K$.
\subsubsection*{Step D}
Use the frequency-angle pairs, $(\omega_m, \theta_k)$, in the evaluation of the objective function.

The grid points $(\bar{\omega}_p, \bar{\theta}_q)$ in {\it Step A} are referred to as {\it virtual sample points}. The technique in~\cite{antoniou2} also allows for increasing the sampling density near the band edges where spikes are more likely to occur. For two dimensions, an analogous approach is to decrease the length and width of the rectangular blocks in {\it Step B} as we approach closer to the edges of the frequency and angular bands. For example, one option to achieve this is to assign fixed frequency-angle points near the frequency and angular band edges by simply setting the lengths and widths of the rectangular blocks to the unit virtual-sampling distance at those locations.
\vspace{-0.1in}
\subsection{Practical Considerations}
To evaluate the parameters that are dependent on the frequency and angles, the 2-D nonuniform variable sampling technique described in Subsection IV-F is used. The 2-D technique results in a complexity reduction by more than an order of magnitude, thereby significantly speeding up the optimization algorithm. The weights $W$ for the slack parameter, $\delta_{rlx}$, in (\ref{iterOptProblem4}) should not be too small as this can make the optimization algorithm unstable and prevent it from converging; at the same time, it should also not be too large as this can slow down the convergence process. Typical values of $W$ that have been found to work well range between 500 to 5000.

Though the convergence speed depends on the initialization point, in most cases the optimization algorithms in (\ref{iterOptProblem4})  converge to a good solution within 50 iterations. In some cases, it has been observed that the solution keeps improving with each iteration, but beyond a certain point the degree of improvement is too small to be of practical significance and the optimization can be terminated. Furthermore, it has been observed that during the optimization iterations the objective function may at some point show very small improvement, or even increase for several iterations, before rapidly decreasing again. To ensure that the optimization is not prematurely terminated, the termination condition is decided by monitoring the values of the objective function for the last $L_o$ iterations. If none of the $L_o$ values are less than the minimum of the objective function obtained before the last $L_o$ iterations, the iteration is terminated. In our experiments, $L_o = 5$ has been found to work well.
\vspace{-0.05in}
\section{Experimental Results}
In this section, we provide comparative experimental results to demonstrate the benefits of the proposed method. The comparison experiments are divided into three subsections on the basis of the prescribed group delay; i.e., in the first, second, and third subsections, the prescribed group delay is set to 0, $[(L-1)f_s^{-1}/2]$, and $[(L-1)f_s^{-1}/4]$, respectively. In our experiments, we compare several variants of the proposed method with modified variants of the competing method. For the proposed method we have the following design variants:
\subsubsection{Designs V1-A and V1-C}
These designs corresponds to the first variant in Section III, where the solution is obtained by solving the convex optimization problem in (\ref{optProblem4}). The prescribed group delay, $\tau_d$, is set to 0 and $[(L-1)f_s^{-1}/4]$ for designs V1-A and V1-C, respectively.
\subsubsection{Designs V1-A(Sym) and V1-C(Sym)}
These designs are special cases of designs V1-A and V1-C, respectively, where the prescribed beampattern is symmetric about $\theta = \pi/2$ with the assumption that the positions of the array sensors are symmetric with respect to the array center. The design ensures that the solution has a beampattern that is perfectly symmetrical, which is obtained by solving the convex optimization problem in (\ref{optProblem4}) with the equality constraint in (\ref{symCondt}) included.
\subsubsection{Design V1-B}
This design is used when the prescribed group delay is $[(L-1)f_s^{-1}/2]$ with the assumption that the positions of the array sensors are symmetric with respect to the array center. The solution for this design is obtained by solving the convex optimization problem in (\ref{optProblem4}) with the equality constraint in (\ref{linPhaseConstr}) included. This design results in beamformers with perfectly linear phase.
\subsubsection{Designs V2-A, V2-B, and V2-C}  
These designs correspond to the second variant, described in Subsection IV-D, where the solution is obtained by solving the 2-step iterative optimization problem with prescribed group delay, $\tau_d$, set to 0, $[(L-1)f_s^{-1}/2]$, and $[(L-1)f_s^{-1}/4]$ for designs V2-A, V2-B, and V2-C, respectively. 

The competing beamformer is obtained by considering the beamformer design in~\cite{mabande2}. In that design, the filter coefficients are obtained by solving an optimization problem where the $L_2$ norm of the error of the beamformer response is minimized while constraining the center of the passband to be unity and the WNG to be above a prescribed threshold. However, since the objective in this paper is to design beamformers where the maximum passband ripple and the minimum stopband attenuation are below prescribed levels, using $L_2$-norm optimization is not appropriate; rather, a much better norm is the $L_\infty$ norm~\cite{wsluBook}. Consequently, we modify the method in~\cite{mabande2} so that the $L_\infty$ norm of the passband error is minimized under the constraint that the $L_\infty$ norm of the stopband error are below prescribed levels. In~\cite{mabande2}, the center of the passband was constrained to unity and the desired passband response was set to unity, thereby constraining the group delay of the beamformer to be 0; in this modification, we generalize the beamformer to have any prescribed group delay by constraining the center of the passband to $B_d(\omega_m, \theta_d)$ and setting the desired passband response to $B_d(\omega_m, \theta_p)$, which is defined in (\ref{bfLinResp}). With these modifications, the modified convex optimization problem is given by
\begin{eqnarray}
\mbox{minimize } & & \| \mathbf{U}_{pb}\mathbf{x} - \mathbf{d}_{pb}^{(lin)} \|_\infty \label{optProblem6}   \\
\mbox{subject to: } & & \| \mathbf{U}_{sb}\mathbf{x} \|_\infty \leq \Gamma_{sb} \notag \\
& & \|\mathbf{A}(\omega_m) \mathbf{x}\|_2 \leq \sqrt{\Gamma(\omega_m)} \ \ \forall \  \ \omega_m \in \Omega \notag \\
& & \mathbf{g}(\theta_d, \omega_m)^T \mathbf{x} = B_d(\omega_m, \theta_d) \ \ \ \forall \  \ \omega_m \in \Omega \notag
\end{eqnarray}
where $\mathbf{x}$ is the optimization variable and $\mathbf{d}_{pb}^{(lin)}$ is defined in (\ref{optProblem4}). Consequently, the optimization problem in (\ref{optProblem6}) can also be combined with the symmetry constraints in (\ref{symCondt}) and (\ref{linPhaseConstr}) to give the following variants:
\subsubsection*{Design C-A}
Here the beamformers are designed using the optimization problem in (\ref{optProblem6}) with the prescribed group delay, $\tau_d$, set to 0.
\subsubsection*{Design C-A(Sym)}
This is a special case of design C-A where the prescribed beampattern is symmetric about $\theta = \pi/2$. This beamformer is obtained by solving the optimization problem in (\ref{optProblem6}) with the equality constraints in (\ref{symCondt}) included.
\subsubsection*{Design C-B}
This design is used when the prescribed group delay is $[(L-1)f_s^{-1}/2]$. This beamformer is obtained by solving the optimization problem in (\ref{optProblem6}) with the equality constraints in (\ref{linPhaseConstr}) included. This design results in beamformers with perfect linear phase.

In both of the subsections, we consider design specifications where the beampattern is symmetric and non-symmetric about $\theta = \pi/2$. For the symmetric case the desired steering angle $\theta_d$, which is used in (\ref{wng}), is set to $\pi/2$ and for the nonsymmetric case to $2\pi/3$. In all designs the WNG is constrained to be above 0 dB, and therefore $\Gamma_{wng} = 1$.

For the iterative optimization problem in (\ref{iterOptProblem4}), $W$ is set to 1000, $\gamma_{small}$ to 0.001, and $\Gamma_\delta (k)$ is defined as
\begin{equation}
\Gamma_\delta (k) = \begin{cases} \gamma_k & k < 20 \\
0.001 & k \geq 20
\end{cases}
\end{equation}
where
\begin{equation}
\gamma_k = \gamma_1 - \frac{(\gamma_1 - \gamma_{19})(k-1)}{20-1}
\end{equation}
$\gamma_1 = 0.5$ and $\gamma_{19} = 0.001$. The speed of sound, $c$, is assumed to be 340 m/s while the sampling frequency, $f_s$, is assumed to be 8 kHz. The frequency and angle dependent parameters for designs V2-A and V2-C are evaluated using the 2-D nonuniform variable sampling technique described in Subsection IV-F with the number of  virtual sampling points for the two dimensions, $P$ and $Q$, set to 200 and 500, respectively, and the number of actual sampling points, $M$ and $K$, set to 22 and 52, respectively. The number of fixed sampling points at each band edge for both the angular passband and stopband parameters are set to 3 and correspond to the last 3 virtual sampling points at the edges; likewise, the number of fixed sampling point at the edges of the frequency band are also set to 3 and correspond to the last 3 virtual sampling points at the edges.

The nonuniform sampling technique is applicable only if the optimization problem in iterative. Hence, for designs that are based on solving a non-iterative convex optimization problem, uniform sampling is used instead. Consequently, for all variants of designs V1 and C, the parameters are evaluated by uniformly sampling the frequency and angular bands and setting the number of sampling points $M$ and $K$ along the two dimensions to each have a value of 200. 

The beamformer performance is evaluated using the following parameters: 
\subsubsection*{Maximum passband ripple}
The parameter is defined as
\begin{equation}
A_p = 20 \log \frac{M_{max}^{(p)}}{M_{min}^{(p)}}
\end{equation}
where
\begin{equation}
M_{max}^{(p)} = \max_{\omega \in \Omega, \theta \in \Theta_p} |B(\omega, \theta)|
\end{equation}
\begin{equation}
M_{min}^{(p)} = \min_{\omega \in \Omega, \theta \in \Theta_p} |B(\omega, \theta)|
\end{equation}

\subsubsection*{Minimum stopband attenuation}
The minimum stopband attenuation is defined as the negative of the maximum stopband gain, given by
\begin{equation}
A_a = -20 \log M_{max}^{(a)}
\end{equation}
where
\begin{equation}
M_{max}^{(a)} = \max_{\omega \in \Omega, \theta \in \Theta_s} |B(\omega, \theta)|
\end{equation}
\subsubsection*{Passband average group delay}
As in~\cite{andreasBook} for the design of digital filters, the average group delay is evaluated by taking the average of the maximum and minimum group delay in the passband, given by
\begin{equation}
\tau_{avg} = \frac{\tau_{min}+\tau_{max}}{2} 
\label{tauAvg}
\end{equation}
where
\begin{equation}
\tau_{min} = \min_{\omega \in \Omega, \theta \in \Theta_p} \tau(\omega, \theta)
\end{equation}
\begin{equation}
\tau_{max} = \max_{\omega \in \Omega, \theta \in \Theta_p} \tau(\omega, \theta)
\end{equation}
and $\tau(\omega, \theta)$ is defined in (\ref{tauDefn}).
\subsubsection*{Passband group delay deviation}
The passband group delay deviation is given by
\begin{equation}
\sigma_{\tau} = \tau_{max} - \tau_{min}
\end{equation}
Alternatively, $\sigma_{\tau}$ can also be expressed as
\begin{equation}
\sigma_\tau = \max_{\theta \in \Theta_p} \sigma_\tau (\theta)
\end{equation}
where
\begin{eqnarray}
\sigma_\tau (\theta) & = & \tau_{max}(\theta) - \tau_{min} \\
\tau_{max}(\theta) & = & \max_{\omega \in \Omega} \tau(\omega, \theta)
\end{eqnarray}
Parameter $\sigma_\tau (\theta)$ will be used in the comparison plots of the group-delay deviation between the various methods in the experiments.
\subsubsection*{Cost function at the optimum}
For beamformers that are designed by solving the convex optimization problem, namely, all variants of designs V1 and C, the cost function at the solution is given by
\begin{equation}
J_{sol} = \| \mathbf{U}_{pb}\mathbf{x}_{sol} - \mathbf{d}_{pb} \|_\infty
\end{equation}
and for beamformers that are designed by solving the iterative optimization problem, namely, design V2, the cost function is given by 
\begin{equation}
J_{sol} = \| |\mathbf{U}_{pb}\mathbf{x}_{sol}|^2 - |\mathbf{d}_{pb}|^2 \|_\infty
\end{equation}
where $\mathbf{x}_{sol}$ is the solution for each of the designs. 
%Note that in Design V2-B (Sym), the phase of the beamformer matches the desired response
\subsection{Examples 1 and 2}
In this subsection, we consider the design of beamformers that have a passband group delay of approximately zero. We compare their performance by observing the design that results in the smallest passband group-delay deviation while ensuring that the maximum passband ripple and minimum stopband attenuation are at similar levels. For the comparisons, we consider the proposed designs V1-A, V1-A(Sym), and V2-A and compare their performances with designs C-A and C-A(Sym), which are $L_\infty$-norm modifications of the method in~\cite{mabande2}.

We consider two beamformer design examples. In the first example the beamformer is symmetric about $\theta = \pi/2$ while in the second example it is non-symmetric. The design specifications for the two examples are given in Tables I and III. Since for the first example the desired beampattern is symmetric, we also include designs V1-A(Sym) and C-A(Sym) in the comparison.
\begin{table}
\begin{center}
\caption{Design specifications for beamformers symmetric about $\theta = \pi / 2$ for example 1}
\label{tab_eg1_specs}
{\footnotesize{
\begin{tabular}{||l|l||} \hline \hline
Parameters & Values \\ \hline 
No. of elements of beamformer & 7 \\
Inter-element spacing, (m)& 0.04 \\
%Sampling frequency, $f_s$, (Hz) & 8000 \\
FIR filter length & 20 \\
Passband region, $\Theta_p$, (deg) & $[80^\circ - 100^\circ]$\\  
Stopband region, $\Theta_s$, (deg) & $[0^\circ - 60^\circ] \cup [120^\circ - 180^\circ]$  \\ 
Frequency band, $\Omega$, (Hz) & [1500 - 3500] \\
Maximum passband ripple (dB) & 0.65 \\
Minimum stopband attenuation (dB) & 5.5 \\
Minimum WNG (dB) & 0 \\
Passband group delay & 0 \\
\hline \hline
\end{tabular}
}}
\end{center}
\end{table}

\begin{table}
\begin{center}
\caption{Design Results for example 1 for a symmetric beamformer.} 
\label{tab_eg1_results}
{\footnotesize{
\begin{tabular}{||l|c|c|c||} \hline \hline
Parameters  & V1-A & V2-A & C-A \\ 
\hline 
Max PB ripple $A_p$, dB  & 0.612 & 0.612 & 0.612 \\
Min SB atten. $A_a$, dB  & 6 & 5.99 & 5.96  \\
$\tau_{avg}$, samples  & -0.088 & -0.00034 &  -0.189 \\
CF value at soln., $J_{sol}$ & 0.03521-$\epsilon$ & 0.07 & 0.0676 \\
$\sigma_{\tau}$, samples  & 0.598 & {\bf 0.0036} & 0.853 \\
\hline 
Parameters  & V1-A(Sym) & V2-A & C-A(Sym) \\ 
\hline 
Max PB ripple $A_p$, dB  & 0.612 & 0.612 & 0.612 \\
Min SB atten. $A_a$, dB  & 6 & 5.99 & 6  \\
$\tau_{avg}$, samples  & -0.033 & -0.00034 &  0.085 \\
CF value at soln., $J_{sol}$ & 0.03521+$\epsilon$ & 0.07 & 0.0679 \\
$\sigma_{\tau}$, samples  & 0.248 & {\bf 0.0036} & 0.295 \\
\hline \hline 
\end{tabular} 
}}
\\ PB: passband; SB: stopband; CF: cost function; $\epsilon = 3 \times 10^{-13}$ \\

\end{center}
\end{table}

\begin{table}
\begin{center}
\caption{Design specifications for a non-symmetric beamformer for example 2}
\label{tab_eg2_specs}
{\footnotesize{
\begin{tabular}{||l|l||} \hline \hline
Parameters & Values \\ \hline 
No. of elements of beamformer & 7 \\
Inter-element spacing, (m)& 0.04 \\
%Sampling frequency, $f_s$, (Hz) & 8000 \\
FIR filter length & 20 \\
Passband region, $\Theta_p$, (deg) & $[110^\circ - 130^\circ]$ \\  
Stopband region, $\Theta_s$, (deg) & $[0^\circ - 90^\circ] \cup [150^\circ - 180^\circ]$  \\ 
Frequency band, $\Omega$, (Hz) & [1500 - 3500] \\
Maximum passband ripple (dB) & 0.70 \\
Minimum stopband attenuation (dB) & 5.5 \\
Minimum WNG (dB) & 0 \\
Passband group delay & 0 \\
\hline \hline
\end{tabular}
}}
\end{center}
\end{table}

\begin{figure}
\begin{center}
\includegraphics[width=0.45\textwidth]{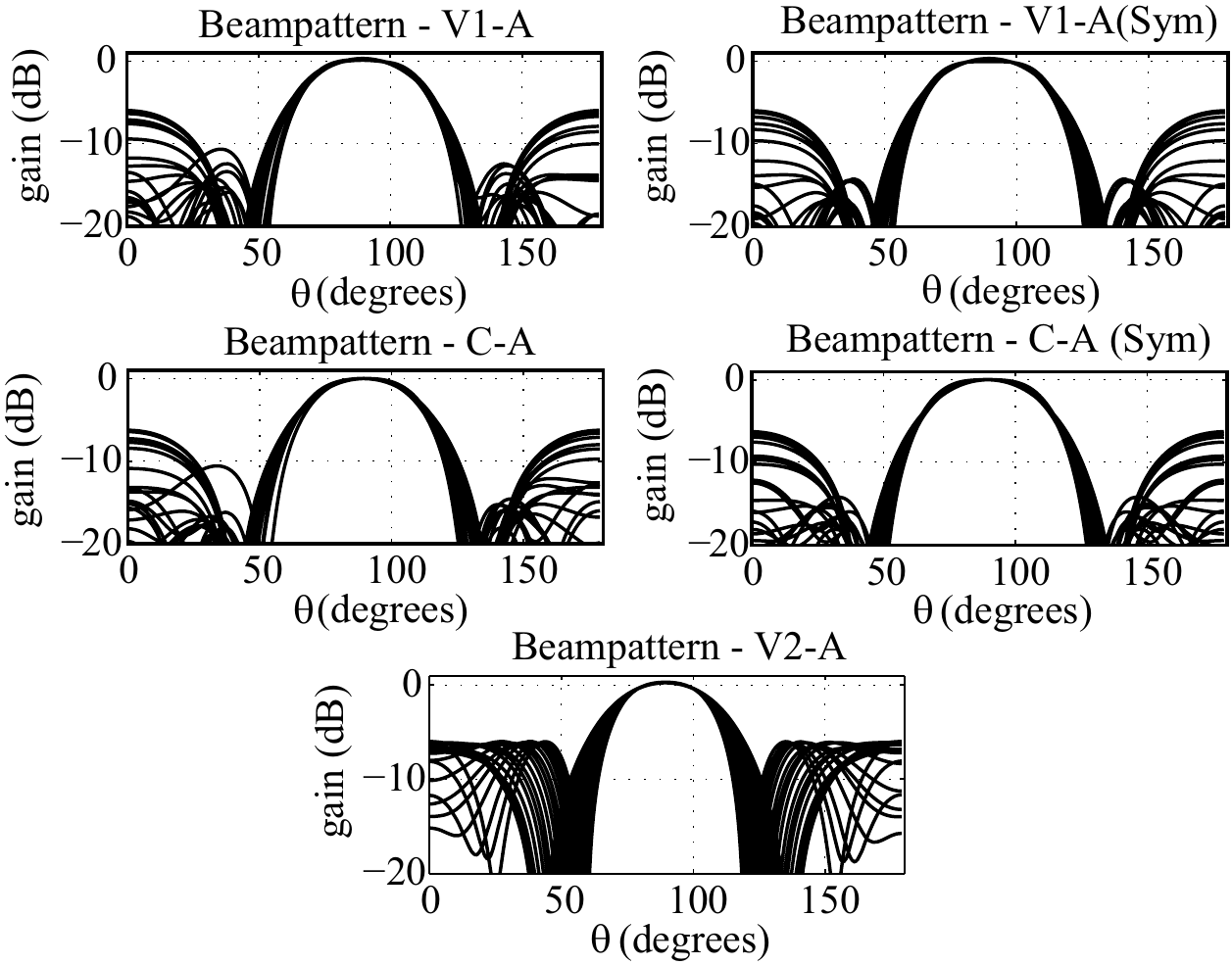}
\caption{Beampattern plots for the various designs for example 1. The plots are obtained by plotting the responses across 20 uniformly sampled frequency-points in the frequency band. }
\label{eg1a}
\end{center}
\end{figure}

\begin{figure}
\begin{center}
\includegraphics[width=0.45\textwidth]{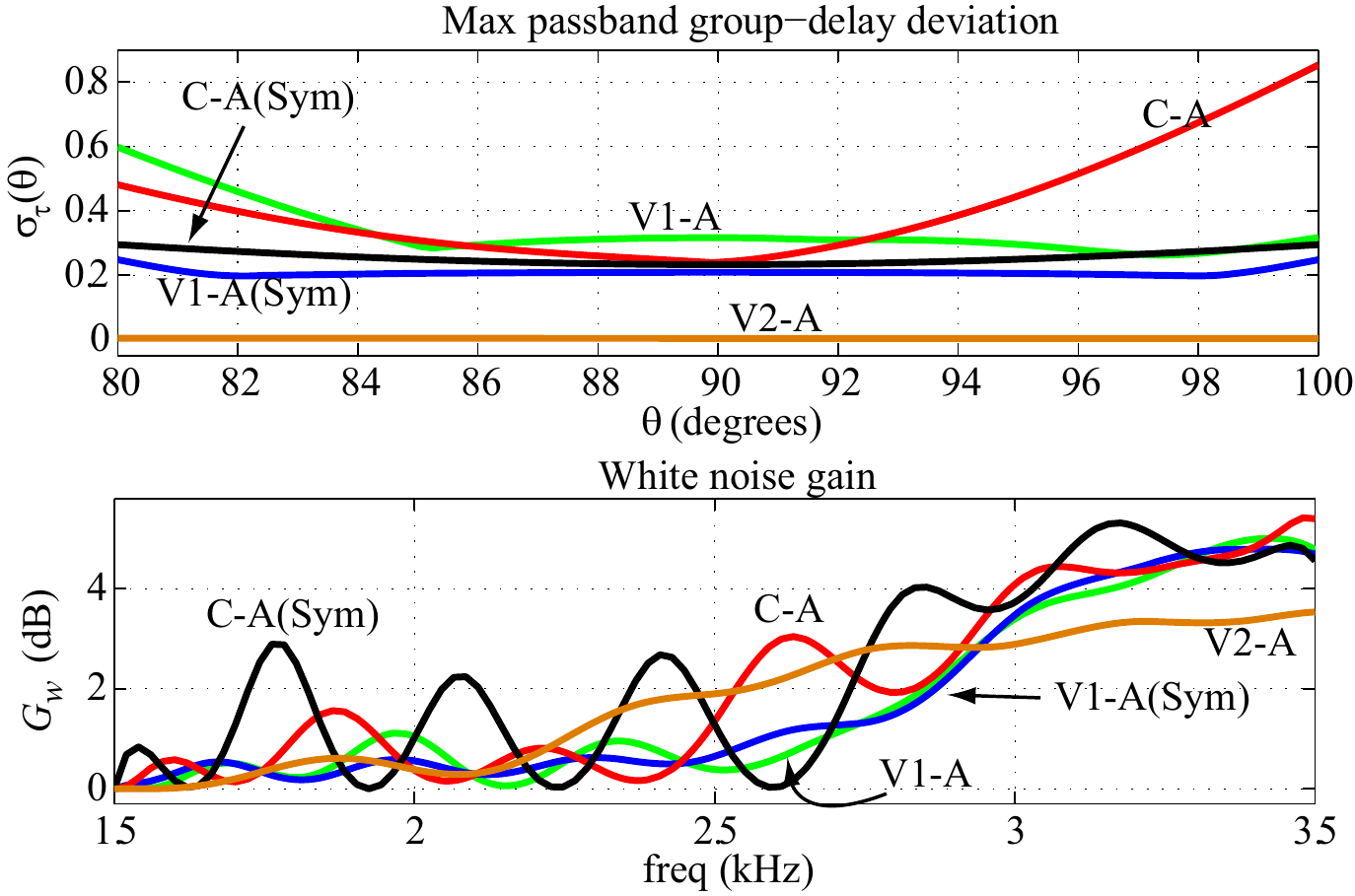}
\caption{Plots of the maximum passband group-delay deviation and white noise gain for the various designs for example 1.}
\label{eg1b}
\end{center}
\end{figure}

\begin{table}
\begin{center}
\caption{Design Results for example 2 for a non-symmetric beamformer}
\label{tab_eg2_results}
{\footnotesize{
\begin{tabular}{||l|c|c|c||} \hline \hline
Parameters  & V1-A & V2-A & C-A \\ 
\hline 
Max PB ripple $A_p$, dB  & 0.674 & 0.672 & NF \\
Min SB atten. $A_a$, dB  & 6 & 6.01 & NF  \\
$\tau_{avg}$, samples  & 0.391 & 0.0001 &  NF \\
$\sigma_{\tau}$, samples  & 1.125 & {\bf 0.0166} & NF \\
\hline \hline 
\end{tabular} 
}}
\\ \hspace{-0.2in} PB: passband; SB: stopband; NF: not feasible
\end{center}
\end{table}

\begin{figure}
\begin{center}
\includegraphics[width=0.45\textwidth]{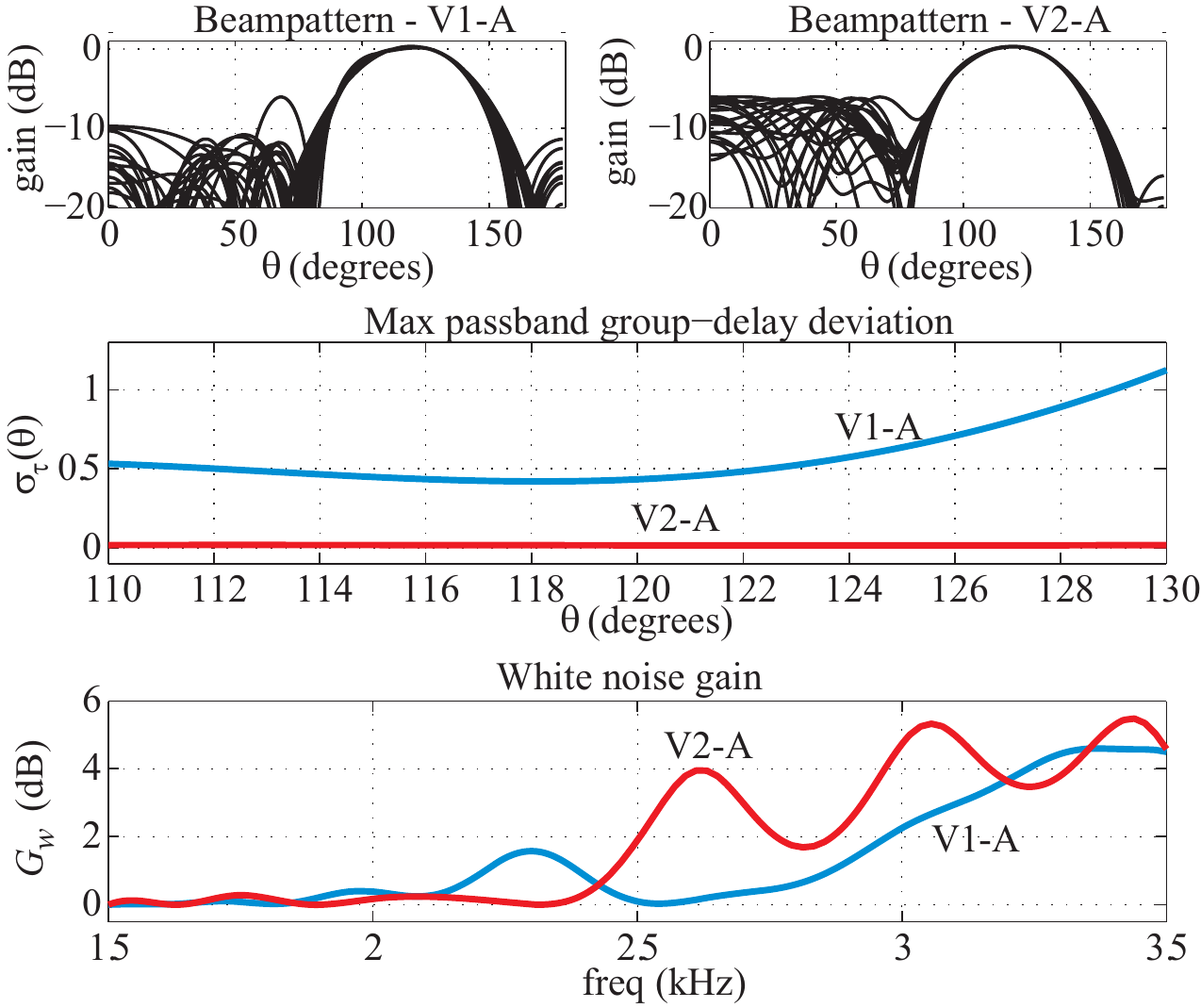}
\caption{Plots of the maximum passband group-delay deviation, beamformer response, and white noise gain for the various designs for example 2. The beampattern plots are obtained by plotting the responses across 20 uniformly sampled frequency-points in the frequency band.}
\label{eg2}
\end{center}
\end{figure}
The comparison results for examples 1 and 2 are summarized in Tables II and IV and the maximum group-delay deviation, beamformer response, and white noise gain are plotted in Figs. 2 and 3, respectively. From Table II we observe that design V2-A has the smallest maximum group delay deviation, $\sigma_\tau$, for similar maximum passband ripple and minimum stopband specifications. Among the non-iterative designs, design V1-A(Sym) has the smallest $\sigma_\tau$. It is interesting to note that both the non-iterative designs with symmetry constraints, namely, designs V1-A(Sym) and C-A(Sym), have smaller $\sigma_\tau$ but slightly larger cost-function values at their optimum, than their counterparts without symmetry constraints. In addition, the designs with symmetry constraints also have perfectly symmetric beampatterns as seen from the plots in Fig. 2.

From Table IV, we again observe that design V2-A has the smallest $\sigma_\tau$ for the non-symmetric beamformer specification of example 2. We also observe that for this example, design C-A does not result in a feasible solution; a possible reason for this infeasibility is the equality constraint in (\ref{optProblem6}), which severely restricts the degrees of freedom in the optimization. 

From the results in this subsection, we can conclude that for the design of beamformers where the prescribed group delay is 0, design V2-A will give the smallest $\sigma_\tau$ for the same values of maximum passband ripple and minimum stopband attenuation. 
\vspace{-0.1in}
\subsection{Examples 3 and 4}
In this subsection, the design specifications for the symmetric and non-symmetric beamformers are the same as in the previous subsection except for two changes: the minimum stopband attenuation is increased to 9.5 dB from 5.5 dB, and the prescribed group delay is set to $[(L-1)f_s^{-1}/2]$. For both of the beamformer designs in examples 3 and 4, we consider the proposed design V1-B and compare its performances with competing design C-B.

For example 3, we observe that both of the designs, V1-B and C-B, have zero group-delay deviation or perfectly linear phase; however, among the two, V1-B has better passband ripple and stopband attenuation than C-B. 

For the non-symmetric beamformer in example 4, we observe that design V1-B has perfectly linear phase, while design C-B is infeasible.

Note that in examples 3 and 4 the microphone positions satisfy the symmetry condition in (\ref{symCondt0}) and, therefore, designs V1-B and C-B are perfectly linear phase; however, in applications where the microphone positions do not satisfy (\ref{symCondt0}), design V2-B can be used instead. 
\begin{table}
\begin{center}
\caption{Design specifications for beamformers symmetric about $\theta = \pi/2$ for example 3}
\label{tab_eg3_specs}
{\footnotesize{
\begin{tabular}{||l|l||} \hline \hline
Parameters & Values \\ \hline 
No. of elements of beamformer & 7 \\
Inter-element spacing, (m)& 0.04 \\
%Sampling frequency, $f_s$, (Hz) & 8000 \\
FIR filter length & 20 \\
Passband region, $\Theta_p$, (deg) & $[80^\circ - 100^\circ]$\\  
Stopband region, $\Theta_s$, (deg) & $[0^\circ - 60^\circ] \cup [120^\circ - 180^\circ]$  \\ 
Frequency band, $\Omega$, (Hz) & [1500 - 3500] \\
Maximum passband ripple (dB) & 0.96 \\
Minimum stopband attenuation (dB) & 9.5 \\
Minimum WNG (dB) & 0 \\
Passband group delay (samples) & 9.5 \\
\hline \hline
\end{tabular}
}}
\end{center}
\end{table}

\begin{table}
\begin{center}
\caption{Design Results for example 3 for a symmetric beamformer}
\label{tab_eg3_results}
{\footnotesize{
\begin{tabular}{||l|c|c||} \hline \hline
Parameters  & V1-B & C-B\\ 
\hline 
Max PB ripple $A_p$, dB  & {\bf 0.953} & 0.981 \\
Min SB atten. $A_a$, dB  & {\bf 10} & 9.55  \\
$\tau_{avg}$, samples  & 9.5 & 9.5 \\
CF value at soln., $J_{sol}$ & 0.0549 & 0.104 \\
$\sigma_{\tau}$, samples  & 0 & 0 \\
\hline \hline 
\end{tabular} 
}}
\\ PB: passband; SB: stopband; CF: cost function; 
\end{center}
\end{table}

\begin{table}
\begin{center}
\caption{Design specifications for a non-symmetric beamformer for example 4}
\label{tab_eg4_specs}
{\footnotesize{
\begin{tabular}{||l|l||} \hline \hline
Parameters & Values \\ \hline 
No. of elements of beamformer & 7 \\
Inter-element spacing, (m)& 0.04 \\
%Sampling frequency, $f_s$, (Hz) & 8000 \\
FIR filter length & 20 \\
Passband region, $\Theta_p$, (deg) & $[110^\circ - 130^\circ]$ \\  
Stopband region, $\Theta_s$, (deg) & $[0^\circ - 90^\circ] \cup [150^\circ - 180^\circ]$  \\ 
Frequency band, $\Omega$, (Hz) & [1500 - 3500] \\
Maximum passband ripple (dB) & 0.98 \\
Minimum stopband attenuation (dB) & 9.5 \\
Minimum WNG (dB) & 0 \\
Passband group delay (samples) & 9.5 \\
\hline \hline
\end{tabular}
}}
\end{center}
\end{table}

\begin{figure}
\begin{center}
\includegraphics[width=0.45\textwidth]{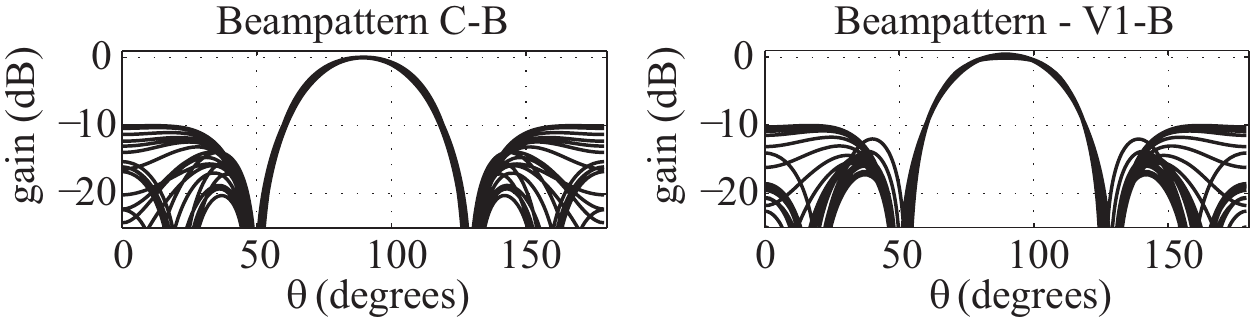}
\caption{Beampattern plots for the two designs for example 3. The plots are obtained by plotting the responses across 20 uniformly sampled frequency-points in the frequency band. }
\label{eg3a}
\end{center}
\end{figure}

\begin{figure}
\begin{center}
\includegraphics[width=0.45\textwidth]{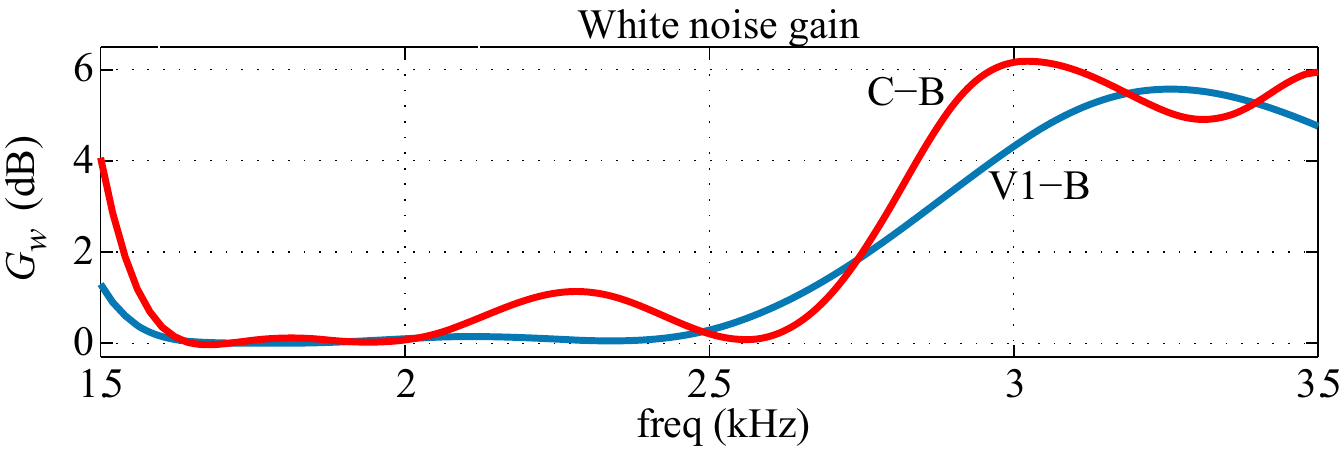}
\caption{Plots of the white noise gain for the two designs for example 3. Both designs in this example have perfectly linear phase and therefore their group-delay deviation is zero.}
\label{eg3b}
\end{center}
\end{figure}

\begin{table}
\begin{center}
\caption{Design Results for example 4 for a non-symmetric beamformer}
\label{tab_eg4_results}
{\footnotesize{
\begin{tabular}{||l|c|c|c||} \hline \hline
Parameters  & V1-B & C-B\\
\hline 
Max PB ripple $A_p$, dB  & 0.977 & NF \\
Min SB atten. $A_a$, dB  & 10 & NF  \\
$\tau_{avg}$, samples  & 9.5 & NF \\
$\sigma_{\tau}$, samples  & 0 & NF \\
\hline \hline 
\end{tabular} 
}}
\\ \hspace{-0.2in} PB: passband; SB: stopband; NF: not feasible
\end{center}
\end{table}

\begin{figure}
\begin{center}
\includegraphics[width=0.45\textwidth]{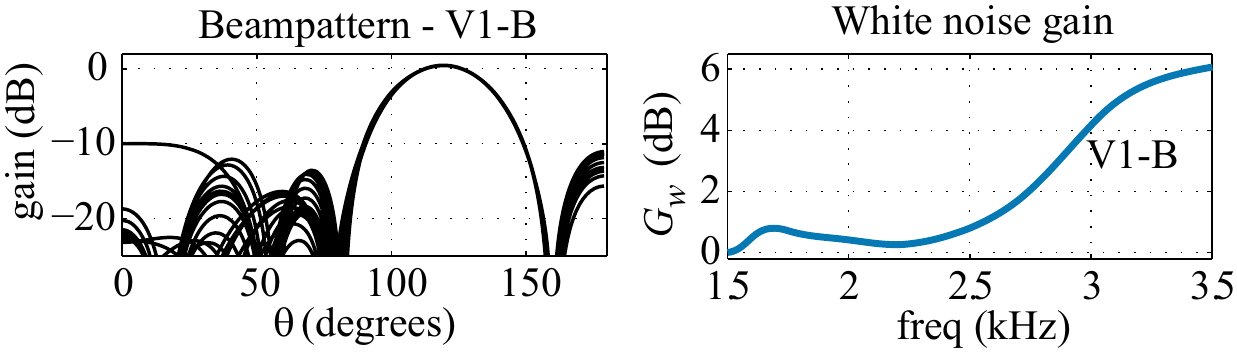}
\caption{Plots of the beampattern and white noise gain for design V1-B in example 4.}
\label{eg4a}
\end{center}
\end{figure}

\subsection{Examples 5, 6 and 7}
We also carried out comparisons for both the symmetric and non-symmetric beamformer design cases when the prescribed group delay is set to $[(L-1)f_s^{-1}/4]$. For the comparisons, we consider designs V1-C and  V2-C for both the symmetric and non-symmetric beamformer cases in examples 5 and 6, respectively. Design C is not considered in these examples as is has been shown in examples 1 to 4 to give beamformers with lower performance than the proposed designs. The design specifications, comparison tables, and plots of the maximum group-delay deviation, beamformer response, and white noise gain for these two examples are given in~\cite{website}. From the results, we find that design V2-C results in much smaller group-delay deviation than V1-C for similar passband ripple and stopband attenuation values. 

In example 7, we compare the original method in~\cite{mabande2} with our proposed method for the symmetric beamformer case. Since the method in~\cite{mabande2} has a prescribed group delay of 0, we use designs V1-A(Sym) and V2-A(Sym) for the comparison. The design specifications, comparison tables, and plots of the maximum group-delay deviation, beamformer response, and white noise gain for this examples are given in~\cite{website}. From the result, we find that design V2-A has the smallest value of $\sigma_\tau$. Furthermore we also observe that though the beamformer designed using the method in~\cite{mabande2} has a slightly better $\sigma_\tau$ than design V1-A, the former has a much larger maximum passband-ripple value, which does not satisfy the given specifications in Table XIII.

The above design examples have shown that the proposed design method yields robust broadband beamformers that are almost linear-phase. Though under certain conditions, the first variant can give beamformers with perfectly linear phase, the second variant is quite general and can be used for obtaining almost-linear-phase beamformers without any restriction on the array configuration or prescribed group delay. In the future, we plan to evaluate the performance of beamformers derived using the proposed techniques using speech and audio signals.

The optimization problems in the examples were solved on a computer running an Intel Core i7-640LM processor using the SeDuMi optimization toolbox for MATLAB. For the first variant, which uses the convex optimization problem in (\ref{optProblem4}), the optimization problem takes anywhere between 2 to 10 minutes to compute. For the second variant, each iteration of the optimization problem in (\ref{iterOptProblem4}) takes less than a minute to compute and the optimization usually converges to a good solution in less than 25 minutes.

\section{Conclusions}
A new method for the design of linear-phase robust far-field broadband audio beamformers using constrained optimization has been described. In the method, the maximum passband ripple and minimum stopband attenuation are ensured to be within prescribed levels, while at the same time maintaining a good linear-phase characteristic at a prescribed group delay in the passband. Since the beamformer is intended primarily for small-sized microphone arrays where the microphone spacing is small relative to the wavelength at low frequencies, the beamformer can become highly sensitive to spatial white noise and array imperfections if a direct minimization of the error is performed. Therefore, to limit the sensitivity of the beamformer, the optimization was carried out by constraining a sensitivity parameter, namely, the white noise gain (WNG) to be above prescribed levels across the frequency band. 

Two novel design variants have been developed. The first variant was formulated as a convex optimization problem where passband error is minimized, while the second variant was formulated as an iterative optimization problem with the advantage of significantly improving the linear phase characteristics of the beamformer under any prescribed group delay or linear-array configuration. In the second variant, the passband group-delay deviation was minimized while ensuring that the maximum passband ripple and stopband attenuation are within prescribed levels. To reduce the computational effort in carrying out the optimization, a nonuniform sampling approach over the frequency and angular dimensions was used to compute the required parameters. Experimental results showed that beamformers designed using the proposed methods have much smaller passband group-delay deviation for similar passband ripple and stopband attenuation than a modified version of an existing method. 
\section*{Acknowledgment}
The authors are grateful to the Natural Sciences and Engineering Research Council of Canada for supporting this work. They also wish to thank the reviewers for their valuable comments and suggestions.
\section*{Appendix}
\subsection{Simplification of the beamformer response due to the symmetry constraints}
For a beamformer satisfying the symmetry conditions in (\ref{symCondt0}) and (\ref{symCondt}), and assuming $N$ to be even for simplicity, the response in (\ref{bmResp}) can be simplified as
\begin{equation}
\bar{B}(\omega, \theta) = \sum_{n=0}^{N/2-1} \sum_{l=0}^{L-1} \bar{x}_{n, l} \bar{g}_{n, l}(\omega, \theta) = \mathbf{\bar{g}}(\omega, \theta)^T \mathbf{\bar{x}}
\label{bmRespX}
\end{equation}
where $\mathbf{x} \in \mathbf{R}^{LN/2}$ and
\begin{eqnarray}
\mathbf{\bar{x}}^T & = & \left[\mathbf{\bar{x}}_0^T ~\mathbf{\bar{x}}_1^T \cdots ~\mathbf{\bar{x}}_{N/2-1}^T \right] \notag \\
 \mathbf{\bar{g}}(\omega, \theta)^T & = & \left[ \mathbf{\bar{g}}_0(\omega, \theta)^T ~\mathbf{\bar{g}}_1(\omega, \theta)^T \cdots ~\mathbf{\bar{g}}_{N/2-1}(\omega, \theta)^T \right] \notag \\
\mathbf{\bar{x}}_n & = & \left[ \bar{x}_{n, 0} ~\bar{x}_{n, 1} \cdots ~\bar{x}_{n, L-1} \right]^T \notag \\
\mathbf{\bar{g}}_n(\omega, \theta) & = & \left[ \bar{g}_{n, 0}(\omega, \theta) ~\bar{g}_{n, 1}(\omega, \theta) \cdots ~\bar{g}_{n, L-1}(\omega, \theta) \right]^T \notag \\
\bar{g}_{n, l}(\omega, \theta) & = & 2 \cos \left(\frac{\omega f_s d_n \cos\theta}{c}\right) \exp (-j\omega l)  \notag
\end{eqnarray}
Note that this simplification has resulted in a reduction of the number of variables in $\mathbf{x}$ from $NL$ to $NL/2$. The simplification for the case where $L$ or $N$ is odd could be derived in a similar manner.
\subsection{Simplification of the beamformer response due to the linear-phase constraints}
For a beamformer satisfying the linear-phase conditions given by (\ref{symCondt0}) and (\ref{linPhaseConstr}), and assuming $L$ and $N$ to be even for simplicity, the response in (\ref{bmResp}) can be simplified as
\begin{equation}
\hat{B}(\omega, \theta) = \sum_{n=0}^{N/2-1} \sum_{l=0}^{L/2-1} \hat{x}_{n, l} \hat{g}_{n, l}(\omega, \theta) = \mathbf{\hat{g}}(\omega, \theta)^T \mathbf{\hat{x}}
\label{bmRespXX}
\end{equation}
where $\mathbf{x} \in \mathbf{R}^{LN/2}$ and
\begin{eqnarray}
\mathbf{\hat{x}}^T & = & \left[\mathbf{\hat{x}}_0^T ~\mathbf{\hat{x}}_1^T \cdots ~\mathbf{\hat{x}}_{N/2-1}^T \right] \notag \\
 \mathbf{\hat{g}}(\omega, \theta)^T & = & \left[ \mathbf{\hat{g}}_0(\omega, \theta)^T ~\mathbf{\hat{g}}_1(\omega, \theta)^T \cdots ~\mathbf{\hat{g}}_{N/2-1}(\omega, \theta)^T \right] \notag \\
\mathbf{\hat{x}}_n & = & \left[ \hat{x}_{n, 0} ~\hat{x}_{n, 1} \cdots ~\hat{x}_{n, L-1} \right]^T \notag \\
\mathbf{\hat{g}}_n(\omega, \theta) & = & \left[ \hat{g}_{n, 0}(\omega, \theta) ~\hat{g}_{n, 1}(\omega, \theta) \cdots ~\hat{g}_{n, L-1}(\omega, \theta) \right]^T \notag \\
\hat{g}_{n, l}(\omega, \theta) & = & 2 \cos \left[\omega \left(\frac{f_s d_n \cos\theta}{c} - \frac{L-1}{2} + l \right)\right] \times \notag \\
& & \hspace{1.5in} \exp \left( -j\omega\frac{L-1}{2} \right) \notag 
\end{eqnarray}
This simplification has resulted in a reduction of the number of variables in $\mathbf{x}$ from $NL$ to $NL/2$.
\subsection{Group delay of a beamformer}
From (\ref{tauDefn}), the group delay of the beamformer can be further expressed as
\begin{equation}
\tau = -\frac{d\theta_B}{d\omega} = -\frac{d}{d\omega} \left\{ \tan^{-1} \frac{\Im[B(\omega, \theta)]}{\Re[B(\omega, \theta)]} \right\}
\label{grpDel}
\end{equation}
where
\begin{eqnarray}
\Im[B(\omega, \theta)] & = & \sum_{n=0}^{N-1} \sum_{l=0}^{L-1} x_{n, l} \sin \phi_{n,l}  \\
\Re[B(\omega, \theta)] & = & \sum_{n=0}^{N-1} \sum_{l=0}^{L-1} x_{n, l} \cos \phi_{n,l}  \\
\phi_{n,l} & = & -\omega \left(\frac{f_s d_n \cos\theta}{c} + l \right)
\end{eqnarray}
Simplification of (\ref{grpDel}) leads to the group delay result in (\ref{grpDelSimple}).

\end{document}